\documentclass[prl,twocolumn,letterpaper,showpacs]{revtex4}

\usepackage{graphicx}% Include figure files
\usepackage{dcolumn}% Align table columns on decimal point
\usepackage{bm}% bold math

\begin{document}
\author{Hayk Harutyunyan$^{1}$}
\email[Email:]{hayk@df.unipi.it, achim.hartschuh@cup.uni-muenchen.de}
\author{Tobias Gokus$^2$}
\author{Alexander A. Green$^3$}
\author{Mark C. Hersam$^3$}
\author{Maria Allegrini$^1$}
\author{Achim Hartschuh$^{2}$}

\affiliation{$^1$ Dipartimento di Fisica "E. Fermi", Universit\'{a}
di Pisa and CNISM, Largo Pontecorvo 3, 56127 Pisa, Italy
\\$^2$Department Chemie und Biochemie and CeNS,
Ludwig-Maximilians-Universit\"{a}t M\"{u}nchen, 81377 M\"{u}nchen,
Germany\\
$^3$Department of Materials Science and Engineering, Department of
Chemistry, Northwestern University, Evanston, Illinois 60208-3108,
USA }

\title{Defect Induced Photoluminescence
from Dark Excitonic States in Individual Single-Walled Carbon Nanotubes}

\date{\today}

\begin{abstract}
We show that new low-energy photoluminescence (PL) bands can be created in
semiconducting single-walled carbon nanotubes by intense pulsed
excitation. The new bands are attributed to PL from different nominally dark
excitons that are "brightened" due to defect-induced mixing of states with
different parity and/or spin. Time-resolved PL studies on single nanotubes
reveal a significant reduction of the bright exciton lifetime upon brightening
of the dark excitons. The lowest energy dark state has longer lifetimes and is
not in thermal equilibrium with the bright state. 
\end{abstract}
\pacs{78.47.+p, 78.55.-m, 78.67.Ch}
\maketitle

Owing to their exceptional electronic properties single-walled
carbon nanotubes (SWNTs) are promising candidates for future
nanoelectronic and biosensing applications as well as narrow band
nanoscale emitting and detecting
devices~\cite{A.2008,W.2007,P.2004b,T.2007}.
Excitons are identified to dominate the absorption and PL emission properties
of these 1--dimensional systems~\cite{F.2005a,J.2005b}. Enhanced
electron--electron interactions due to the reduced dimensionality
lead to exceptionally large binding energies of the excitons that
are shown to exist even in metallic SWNTs~\cite{F.2007a}.
Theory predicts a manifold of excitonic bands below the free
electron--hole continuum of semiconducting
SWNTs~\cite{H.2004b,V.2005b,Tretiak2007b,E.}. Besides the optically
active odd parity singlet excitons additional even parity
dipole--forbidden singlet as well as triplet excitons are expected
to occur. Most importantly, some of these bands form non--emissive
states that are lower in energy than the lowest bright state. This
complex sequence of excitonic states and the non--radiative
relaxation channels associated with them presumably have an
important impact on the low PL quantum yield of SWNTs and fast
exciton decay rates~\cite{F.2004,A.2005a,S.2008}.

The direct experimental proof for the existence of dark excitonic
states in SWNTs was presented applying two--photon photoluminescence
excitation spectroscopy~\cite{F.2005a,J.2005b} and measurements of
magnetic brightening of the SWNT PL~\cite {J.2007c}. Low-energy
forbidden states were also used to explain the dynamics observed in
pump--probe experiments~\cite{Z.2007a,H.2006a} and the temperature
dependence of PL intensities~\cite{I.2007}. In addition, recent
results on ensemble~\cite{W.2007a} and individual nanotube
samples~\cite{O.2007a} have shown low energy satellite peaks in the
PL spectra. These peaks were attributed to emission from low lying
dark excitonic states while the mechanisms enabling
optically forbidden transitions and the interplay between
bright and dark excited states remain to be clarified.

In this Letter, we report on the creation of low-energy emission
bands in the PL spectra of individual $(5,4)$ and $(6,4)$
SWNTs upon high power pulsed laser irradiation at room temperature.
The persistence of these bands in subsequent low power measurements
indicates that irreversible distortions of the nanotube structure
efficiently ``brighten'' forbidden states via disorder induced
mixing of excitonic states in agreement with theoretical
predictions~\cite{V.2005b,C.2005a}. The clear distinction between 
additional emissive features belonging to a certain nanotube and PL
bands from other nanotubes is made possible by observing single
nanotube spectra before and after high power irradiation and by
recording the polarization dependence of bright and dark emission
bands. 
While the  decay times of the allowed transition
are in the range of 1 to 40 ps~\cite{T.2008}, far longer dark state
lifetimes of up to 177 ps have been observed. 
Based on the spectroscopic properties of the lowest dark state emission and
its observation upon nanotube exposure to gold that is predicted to create
high local spin densities~\cite{E.2004a,A.2008a}, 
we suggest that low energy emission results from
triplet exciton recombination facilitated by magnetic defects and impurities.

\begin{figure}[ht]
    \includegraphics[width=0.8\columnwidth,keepaspectratio=true,draft=false]{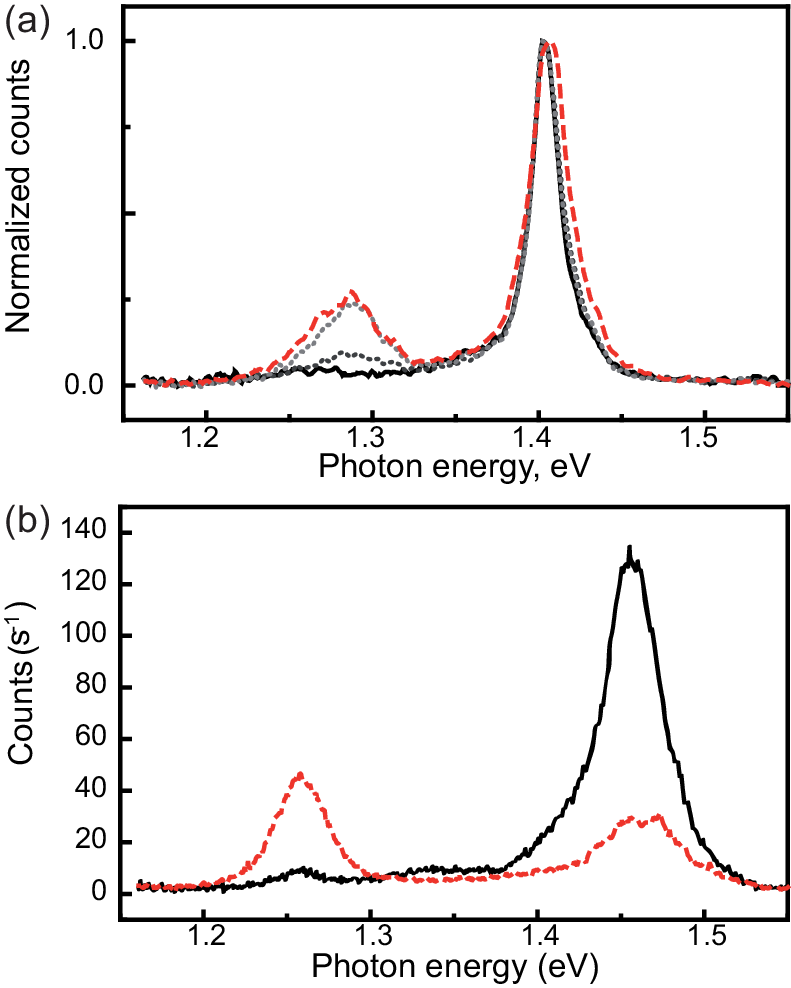}
    \caption{Creation of low energy satellite peaks in the PL spectrum of 
      a $(6,4)$ SWNT (a) and a $(5,4)$ SWNT (b).
       Initial spectra (solid black lines) acquired at low
      excitation intensity $I_0 = 3\cdot 10^{13}\rm photons \cdot pulse^{-1}
      cm^{-2}$ and  
    considerably modified spectra (dashed red lines) of the same nanotubes
    acquired after exposure to high excitation intensity $\sim 17 \cdot I_0$. 
    Low energy satellite contributions shifted by $\sim30-60$ meV (DE$_2$) and
    $\sim 110-190$ meV (DE$_1$) with respect to the bright exciton emission
  are assigned to emission from optically dark states.
    Dotted grey lines in (a) were acquired sequentially between initial and
    final spectrum at intermediate intensity $\sim 7 \cdot I_0$ illustrating
    the step-like creation of the additional bands.
    }
    \label{fig:1}
\end{figure}

Time-correlated single-photon counting (TCSPC) is used in
combination with confocal microscopy to perform single SWNT
spectroscopy and time--resolved PL measurements~\cite{T.2008}. 
Spatially isolated individual SWNTs were obtained by
spin-coating a small volume of micelle-encapsulated CoMoCat material onto a
glass coverslip~\cite{A.2007d}. Laser excitation was provided by a Ti:Saphire
oscillator operated at a photon energy of 1.63 eV and a repetition rate of 76
MHz.
The PL spectra were recorded with a CCD camera
and a fast avalanche photodiode was used to acquire PL transients. 

Fig.~\ref{fig:1} shows the generation of low energy satellite PL bands for two
individual nanotubes. Initial spectra (solid black lines) acquired at low
excitation intensity ($I_0 = 3\cdot 10^{13}\rm photons \cdot pulse^{-1}
      cm^{-2}$) show a single emission peak centered at 1.41 eV (a) and 1.46
      eV (b) assigned to the allowed bright exciton 
(BE) in $(6,4)$ and $(5,4)$ nanotubes, respectively~\cite{S.2002}.
Irradiation of the nanotubes for 10--100 seconds with an order of
magnitude higher pulse intensity $\sim 17 \cdot I_0$  results in some cases in
significantly modified spectra (dashed red lines in Fig.~\ref{fig:1}) with
additional redshifted shoulders and new spectral features
transferring substantial spectral weight to these satellite
peaks. Importantly, no such spectral changes were induced at the corresponding
averaged power levels using continuous--wave (cw) excitation suggesting that
high pulse intensities initiating multi-photon processes
are crucial to induce these modifications. High power cw excitation,
on the other hand, mainly leads to photobleaching and blinking of
nanotube PL~\cite{C.2008}. Satellite peaks for different $(6,4)$ and
(5,4) nanotubes consistently appear at similar energies and can be
roughly divided into two groups with redshifts of $\sim 110-190$ meV
(DE$_1$) and $\sim30-60$ meV (DE$_2$) in good agreement with
Ref.~\cite{O.2007a}. The same energy splittings (130 and 40 meV)
were predicted  for the $(6,4)$ nanotube and attributed to triplet and
even parity singlet excitons, respectively~\cite{E.}.

\begin{figure}[ht]
    \includegraphics[width=0.8\columnwidth,keepaspectratio=true,draft=false]{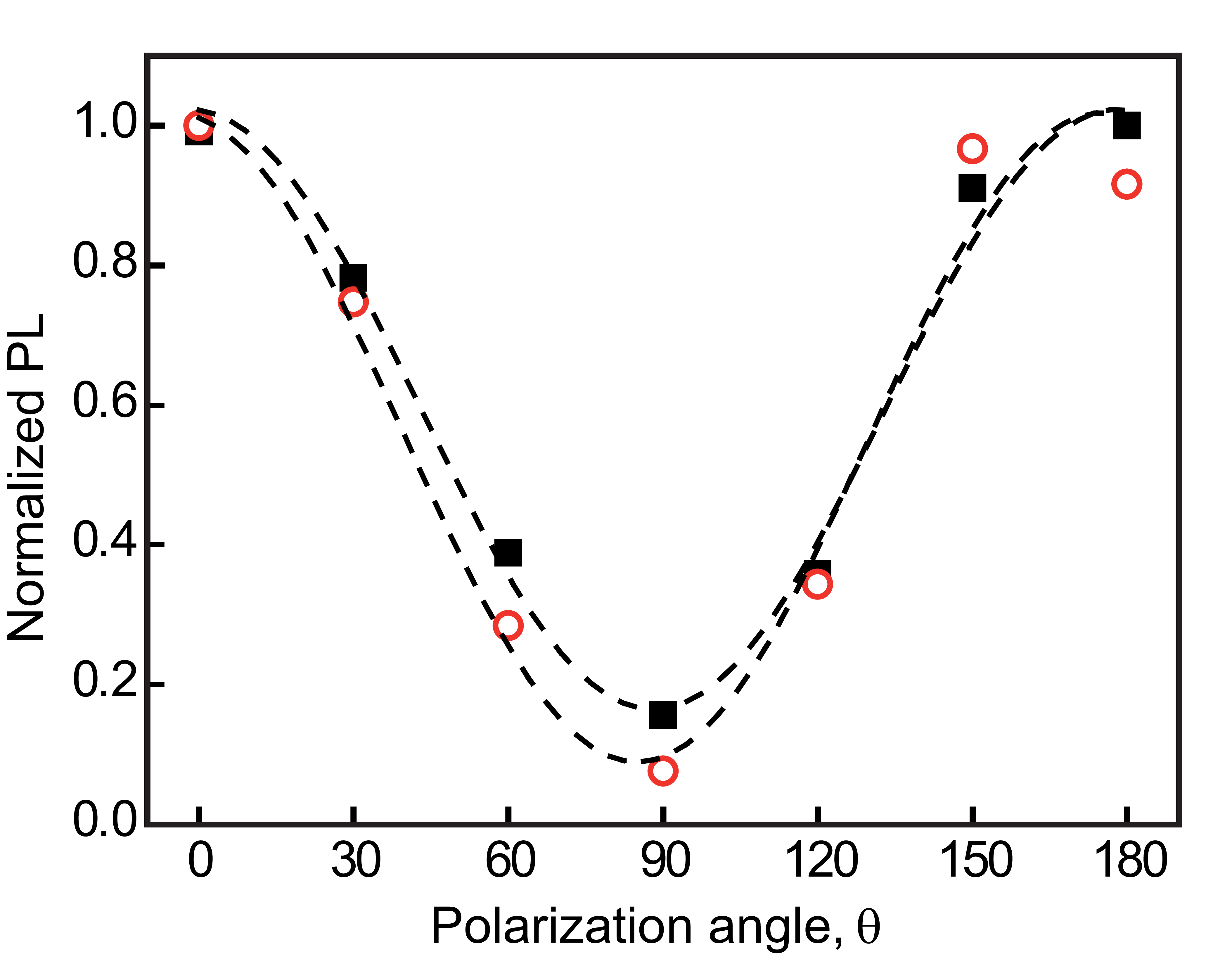}
    \caption{Polarization dependence of the PL emission for the
    bright exciton BE (open circles) and the dark exciton DE$_1$ (filled squares)
    determined from a series of spectra recorded for the same (6,4) nanotube.
    The dashed lines are $cos^{2}\theta$ fits.}
    \label{fig:3}
\end{figure}

The polarization analysis of PL emission of the original BE peak and
the newly created satellite DE$_1$ (Fig.~\ref{fig:3}) show the same
$cos^{2}\theta$ behavior proving that the emission bands belong to
the same nanotube and indicating that the redshifted emission
originates from an intrinsic state of the SWNT. Furthermore, consistent 
appearance of the new bands in DNA wrapped SWNTs and nanotubes embedded 
in PMMA matrix (Fig.~S1 of the supplementary information~\cite{a})
shows that the effect is not due  
to a chemical reaction specific to sodium cholate surfactant.

Based on these experimental observations we conclude that during the
intense irradiation the structure of the nanotube is modified by
creation of local defect sites. These defects alter the local
symmetry of the nanotube partially removing restrictions for the
population and subsequent emission from intrinsic dark
states~\cite{V.2005b}. Although the spectral changes were generally
irreversible, some nanotubes exhibited a 
reversible power dependence of the amplitude of the redshifted peak.
This could indicate that some reversible structural defects are
stable only at high thermal energies of the nanotube. While defects are
probably ubiquitous in SWNTs, 
thereby explaining the observation of multiple PL peaks in
the literature~\cite{J.2004c,W.2007a,O.2007a}, their creation might be
suppressed at low temperatures, where such effects were
not observed under similar experimental conditions~\cite{K.2008}.

\begin{figure}[ht]
    \includegraphics[width=0.7\columnwidth,keepaspectratio=true,draft=false]{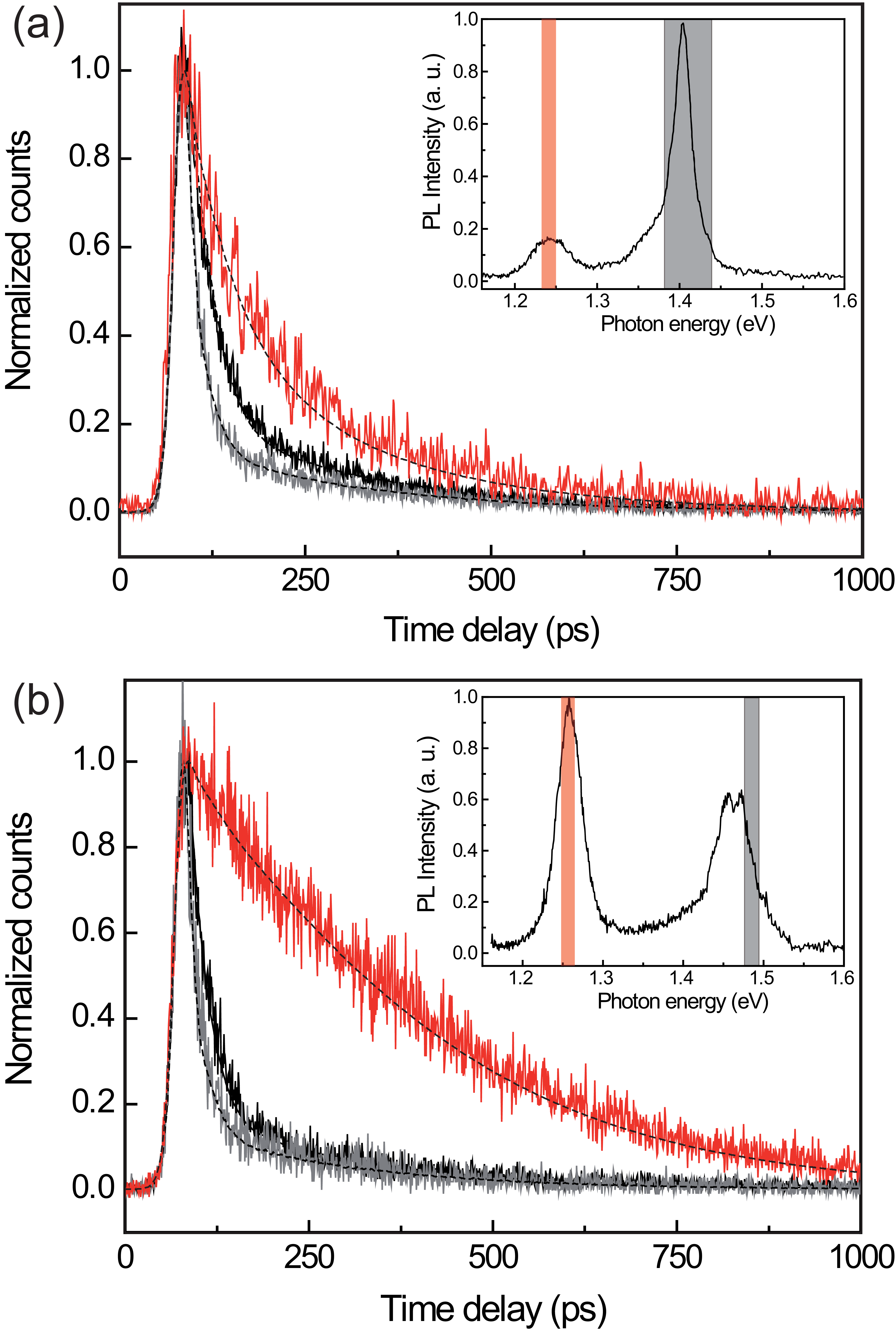}
    \caption{PL transients visualizing the decay dynamics of different
      emission peaks in the spectra of individual SWNT (insets) for two
      different chirality nanotubes: (6,4) (a) and (5,4) (b). Black curves
      show the decay of the BE state detected in the grey shaded
      spectral range in the insets before creation of the low energy
      satellites. After creation of the dark state emission the decay rate of
      the bright exciton  is increased significantly (grey lines).
      The decay of the dark excitonic state DE$_1$, shifted by 110-190 meV
      with respected to the BE, is substantially slower (red
      curves).
      PL decay traces detected for smaller shifts of 30-60 meV resulting from
      DE$_2$ are identical to those of the bright exciton (data not shown).
      Dashed lines are exponential fits to the data (see text).}
    \label{fig:2}
\end{figure}

To determine the population dynamics of both dark and bright
excitonic states and to study the effect of the created disorder we
have performed time--resolved PL measurements of the different
emission bands before and after creation of emission satellites.
Fig.~\ref{fig:2} depicts representative PL transients detected from
the shaded spectral areas (shown in the insets) for two individual
nanotubes of two different chiralities: (6,4) (a) and (5,4) (b). Two
important conclusions can be drawn from this data. First, upon
creation of the satellite peaks the bright exciton lifetime is
decreased (grey curve) compared to the initial decay (black curve),
and second, the DE$_1$ emission  has a much longer decay time.
Monoexponential fits (dashed lines) to these transients give the
lifetimes of the main emission peaks
 before and after creation of the redshifted band of
20 ps and 6 ps for the (6,4) nanotube Fig.~\ref{fig:2}(a) and 13 ps
and 2 ps for the (5,4) tube Fig.~\ref{fig:2}(b), respectively.
Importantly, the emission bands with smaller energy shifts in the
range of several 10 meV (DE$_2$) show exactly the same decay
behavior as the main peak, confirming that these states are in
thermal equilibrium with the bright exciton at room temperature
(data not shown)~\cite{C.2005a,I.2007,G.2007a}.
The decay of dark exciton DE$_1$ is dominated by much longer time
constants, 65 ps and 177 ps for the (6,4) and the (5,4) nanotube in
the present example, as would be expected for a weakly-allowed
transition. Thus, other origins of the low-energy bands such as
phonon replica and bi-excitons can be ruled out based on this slow
decay dynamics. Additionally, we observed a fast decay component (8 ps
and 2 ps) with far smaller photon flux (about factor 1/20) possibly
caused by heterogeneities along the nanotube introduced by the
defects or by a more complicated structure of the DE$_1$ excitonic
manifold. Because of the large separation of the emission peaks and
the detected spectral windows (shaded areas in insets in
Fig.~\ref{fig:2}) overlapping emission contributions from the BE
state are not sufficient to explain this decay component.
On the other hand, the decay dynamics of the dark exciton is
possibly complex since it involves local defects controlling both
initial population and emission by making the forbidden state
weakly allowed and possibly also causing non-radiative quenching.
Measurements on a number of other (6,4) and (5,4) SWNTs
consistently show the same effects with DE$_1$ lifetimes
ranging up to 177 ps. Decay times derived from time--resolved
PL and pump--probe data in the range of 50 - 300 ps with small
relative amplitudes have been reported before~\cite{S.2007d,L.2004}
from ensemble samples as part of multiexponential decay. We
speculate that these decay times could originate from the newly created
states observed here.

Now we discuss the more rapid decay of the bright exciton in the
presence of the redshifted peaks (Fig.~\ref{fig:2}). Since the
amplitude of the BE peak is decreased we conclude that disorder
inducing defects are responsible for additional radiationless
relaxation channels depopulating the bright excitonic state. Two possible
competing channels can be distinguished. First, population transfer
to the dark state DE$_1$ mediated by the introduced defects and
second, decay to the ground state facilitated by enhanced
exciton--phonon coupling due to defect associated local phonon
modes~\cite{B.2007c}. Both relaxation channels require
propagation of the bright exciton along the nanotube to enable
interaction with localized defects, therefore faster decay also
serves as an evidence for the mobility of excitons in
nanotubes~\cite{A.2005a,weisman,C.2008}. 
Population transfer from bright to dark states on the other hand would result
in a delayed rise of the DE$_1$ emission with a rise time equal to the
decay time of the bright state. Such a delayed rise, that would be
 detectable in our measurements especially for nanotubes with
longer decay times of the BE state of up to 25 ps,
was not observed suggesting that a substantial
fraction of the dark state population is built up directly upon
photoexcitation. Importantly, the fact that the bright exciton maintains a
different and finite lifetime in the presence of the dark state DE$_1$
clearly shows that these two states are not
in thermal equilibrium. 

Based on the longer lifetimes of DE$_1$ state and a good agreement of its
emission energy with theoretical predictions~\cite{E.} we speculate that this
PL band is most likely due to triplet state emission. The intersystem crossing
leading to triplet emission can be assisted by coupling to high spin density
states created by sidewall modification of the nanotube such as vacancy
creation~\cite{P.O.2006,P.O.2003}. The energy of about 5 eV needed to create a 
vacancy~\cite{P.O.2006} can be provided through multi-photon excitation
processes explaining the high pulse energies required for the creation of
DE$_1$. In general, magnetic impurities increasing spin-orbit coupling could
also be formed by trace amounts of residual catalyst materials explaining the
observation of dark state emission in other nanotube materials reported in
literature. To test this possibility we treated the SWNTs with a pH
neutralized, aqueous solution of gold~\cite{a} which induces spin polarized
states with significant magnetic moments when adsorbed on SWNT~\cite{E.2004a}
or graphene surface~\cite{A.2008a}. Covering the sample with a gold solution
results in similar changes in the single nanotube PL spectrum without
requiring high power pulsed excitation (Figs.~S2 and S3 of the supplementary
information~\cite{a}). The efficiency of 
brightening of the dark states is especially high when the aqueous SWNTs
solution is premixed with the gold solution before deposition thus facilitating
the surface adsorption of the metal. In these samples the majority of the
$(6,4)$ nanotubes exhibited low energy emission sattelites, indicating that
the same emissive DE$_1$ state is brightened. This has been further confirmed
by time resolved measurements showing a broad distribution of lifetimes in the
range of 7 ps to 150 ps, considerably longer than for the BE
emission. Importantly, no additional PL bands have been observed in
control experiments on single nanotubes deposited on gold films (not
shown) as well as near-field optical experiments using sharp gold 
tips~\cite{hartschuh08} indicating that the new PL bands are not created by
metal surface induced electromagnetic field enhancement. 

In conclusion, we demonstrated that nominally dark excitonic states
in carbon nanotubes can become emissive after exposure to high
excitation intensities and by adsorption of gold. We suggest that local
defects induce mixing of different excitonic states and relaxation of
selection rules via perturbation of the electronic 
structure. Our single nanotube measurements show that the
recombination time of the excitons can be modified by the presence
of disorder and that PL from the same nanotube can have decay rates
varying by 2 orders of magnitude depending on the detected spectral
range.
While these findings are relevant for nanotube photophysics, they also
indicate possible novel routes for the engineering of SWNT optical
properties.

\begin{acknowledgments}
We thank Nicolai Hartmann for valuable experimental help. Financial support by
the DFG through grant HA4405/4-1 and Nanosystem 
Initiative M{\"u}nchen (NIM) is gratefully acknowledged. This work
was also funded by the U.S. National Science Foundation under Award
Numbers EEC-0647560 and DMR-0706067. H. H. acknowledges the financial
support from School of Graduate Studies G. Galilei (University of
Pisa).
\end{acknowledgments}

\end{document}